\documentclass[letterpaper, 10 pt, conference]{ieeeconf}  

\IEEEoverridecommandlockouts                              

\usepackage{graphics} 
\usepackage{epsfig} 

\title{\LARGE \bf
A Magnetically-Triggered Soft Capsule for On-Demand Mucus Collection
}

\author{Xingzhou Du$^{1, 2}$, Kai Fung Chan$^{1, 2}$, Xianfeng Xia$^{2}$, Philip Wai Yan Chiu$^{2, 3, 5}$ and Li Zhang$^{2, 4, 5}$
\thanks{$^{1}$Xingzhou Du and Kai Fung Chan are with the Department of Biomedical Engineering, The Chinese University of Hong Kong, Hong Kong, China.
        }%
\thanks{$^{2}$Xingzhou Du, Kai Fung Chan, Xianfeng Xia, Philip Wai Yan Chiu and Li Zhang are with the Chow Yuk Ho Technology Centre for Innovative Medicine, The Chinese University of Hong Kong, Hong Kong, China.
       }%
\thanks{$^{3}$Philip Wai Yan Chiu is with the Department of Surgery, The Chinese University of Hong Kong, Hong Kong, China.
        }%
\thanks{$^{4}$Li Zhang is with the Department of Mechanical and Automation Engineering, The Chinese University of Hong Kong, Hong Kong, China.
        }%
\thanks{$^{5}$Philip Wai Yan Chiu and Li Zhang are with the T Stone Robotics Institute, The Chinese University of Hong Kong, Hong Kong, China.}
\thanks{Corresponding author to {\tt\small lizhang@mae.cuhk.edu.hk}.}
}

\begin{document}

\maketitle
\thispagestyle{empty}
\pagestyle{empty}

\begin{abstract}
In this work, we present a soft capsule for mucus collection in human intestine for diagnostic purpose, with reduced risk of tissue damage compared with other biopsy methods. The capsule implements passive locomotion and the sampling process is triggered by magnetic field using a permanent magnet, which is placed on the skin above the region of interest (ROI). The capsule contains a soft vacuum chamber which is sealed with wax. When magnetic field and mucus are present simultaneously, the circuit inside the capsule will be closed and nichrome wire will start to generate heat to melt the wax, and mucus will be collected into the vacuum chamber due to air pressure. Experiments on heating capability of nichrome wire, mucus collection and reliability of the capsule are conducted to validate this design.

\end{abstract}

\section{Introduction}

Capsule endoscope is an evolutionary tool for diagnosing diseases in gastrointestinal (GI) tract. Since the first capsule endoscope made by Given Imaging Inc. was approved by FDA in the United States in 2003, millions of patients have benefited from this comfortable endoscopic procedure \cite{c1}. The clinical application of capsule endoscopes makes it possible to inspect the entire small intestinal areas that was hard to reach using traditional endoscopes and provide painless methods for diagnosing GI tract diseases such as obscure gastrointestinal bleeding (OGIB), polyps, Celiac disease, and Crohn's disease \cite{c2}, \cite{c31}. However, there are still limitations for commercialized capsule endoscopes that many practical functions such as drug delivery and biopsy cannot be completed \cite{c30}.

Among the years, many researchers have made various progress in the design of biopsy tools used for capsule endoscopes. In 2013, a biopsy module using magnetic torsion spring mechanism and cylindrical blade was invented by M. Simi et al. \cite{c4}, but the large volume restricted its application in small-sized capsule endoscopes. Then in 2015, V. Le, et al. improved the design and invented a biopsy module using biopsy razer which was actuated by shape memory alloy (SMA) and a pair of small permanent magnet \cite{c5}, and in 2016, V. Le, et al. introduced a compact biopsy structure that based on SMA and cylindrical blade \cite{c6}, however, for these two designs, possible damage to healthy tissue might be caused by inaccurate control of the capsule locomotion. The design has also been improved in 2017, a gripper tool actuated by external magnetic field was developed by V. Le, et al. and the camera was allocated on the lateral side of the capsule, in the same direction with the gripper \cite{c7}. However, the control method of this capsule will be complicated since two magnets with orthogonal directions were implemented. In 2017, based on the magnetically actuated soft capsule endoscopes (MASCE) design \cite{c8}, a fine needle biopsy tool was invented by D. Son, et al. \cite{c9}, but as the capsule was soft, the needle might stretch out by accident due to the movement of stomach, which is dangerous to patients.

\begin{figure}
      \centering
      \includegraphics[scale=.71]{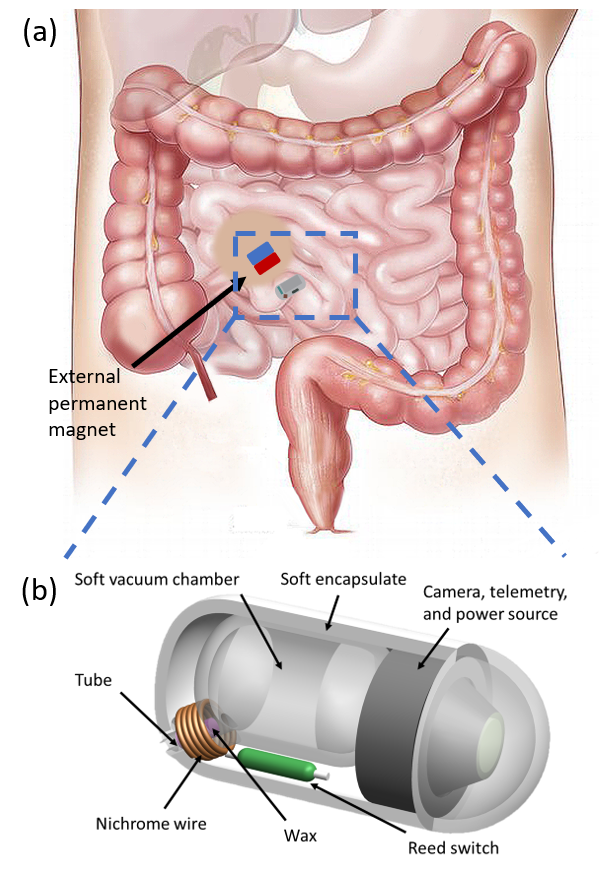}
      \caption{(a) The conceptual working scenario of the soft capsule for on-demand mucus collection. It consists of a permanent magnet that sticks on patient's skin and a soft capsule for mucus collection. (b) Schematics of all the components inside the capsule. }
      \label{scenerio}
\end{figure}

Here we consider implementing biopsy on intestinal mucus instead of tissue to prevent potential damage of the epithelium. Intestinal mucus has significant functions in maintaining normal functions of the digestive system, including receiving pancreatic enzymes (digestion), absorption of nutrients, and protecting epithelium from pathogenic bacteria \cite{c10}. Sampling and analyzing mucus is useful in diagnosing diseases such as malfunction of digestive glands, bacterial or viral infections (e.g. small intestinal bacterial overgrowth), intestinal parasite infections (e.g. ascaris, amebic dysentery, etc.), inflammatory bowel disease (e.g. Crohn's disease), and presence of pathogenic bacteria (e.g. Helicobacter pylori) \cite{c11}-\cite{c15}. Compared with traditional clinical methods, such as fine-needle aspiration that is uncomfortable and possible to damage tissue, aspiration test that is inaccurate, stool sample test that is difficult to separate component, and trial of treatment that may make patients miss optimal therapeutic period, mucus collection using capsules have various advantages including less tissue damage, increased accuracy and decreased discomfort. Besides, capsules have the capability to reach deeper inside small intestine and collect mucus in required location.  

In this paper, we present a soft capsule used for on-demand mucus collection. The capsule consists of a soft vacuum chamber, a reed switch, a nichrome wire, a camera module and a power source. Magnetic field is chosen as a trigger signal for its wide application in miniaturized devices, ranging from millimeter scale \cite{c8}, \cite{c16} all the way down to nanometer scale \cite{c25}-\cite{c27}. The capsule implements passive locomotion that is propelled by natural peristaltic constrains of GI tract. An external permanent magnet attached to the skin above the ROI to mark the sampling position, and when the capsule reaches the location, the magnetic field will close the reed switch and start the mucus collection process. The capsule will then carry the sample outside human body. Further examinations on bacteria, virus, or chemical components could then be conducted based on the sampled mucus.


\section{Design and Operation}

\subsection{Capsule Design}

The working scenario of the mucus collecting capsule and the main components are depicted as Fig. 1. A permanent magnet is attached to patient's skin above the sampling location in small intestine, to trigger the capsule when it moves to the region of interest (ROI) passively. The capsule contains a soft chamber, which could be pressed to generate a proximate vacuum environment inside. A tube is designed to connect the chamber to outside environment and could be tightly blocked with wax after the chamber is loaded by vacuum. Around the tube twines a nichrome wire that is used to generate heat and melt the wax. A reed switch is used to sense the magnetic field of specific strength. A camera module and a power source are integrated in the capsule for viewing and provide power, respectively. The nichrome wire, reed switch and power source make up a heat generation circuit, where nichrome wire transfers electricity to heat and reed switch is in charge of close the circuit under magnetic field.  

\begin{figure}
      \centering
      \includegraphics[scale=.8]{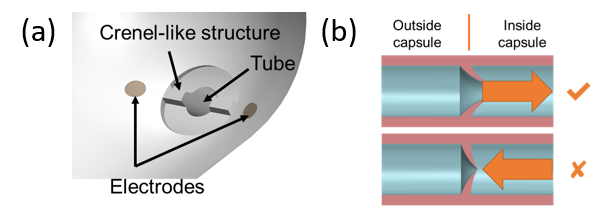}
      \caption{(a) Crane-like structure is designed to prevent capsule sticking to tissue when collecting mucus, and electrodes are used to detect mucus outside the tube, and (b) structure inside tube is designed to prevent leakage and protect collected sample.}
      \label{detail}
\end{figure}

\begin{figure}[thpb]
      \centering
      \includegraphics[scale=.45]{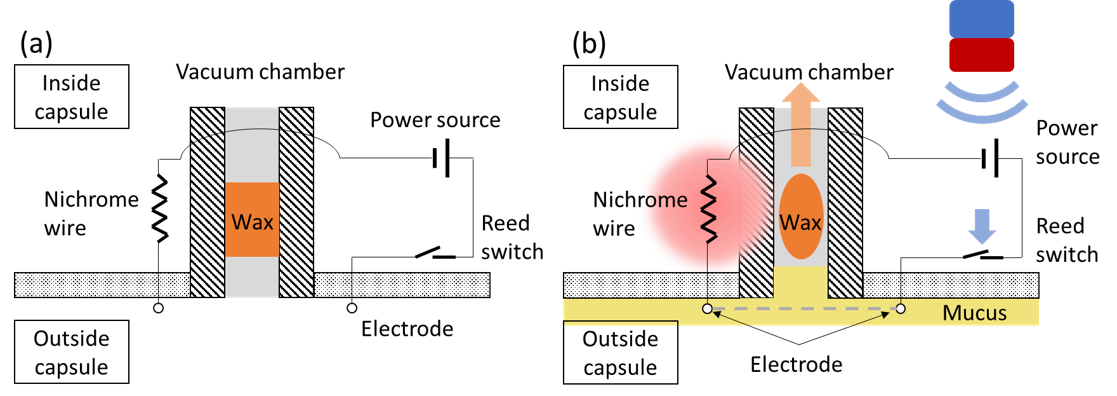}
      \caption{(a) Before triggered by magnetic field, the tube is sealed with wax to keep the vacuum environment inside soft chamber, the heating circuit is not working. (b) When magnetic field and mucus present at the same time, the circuit is closed and nichrome wire melts the wax. Then the mucus will be collected due to air pressure.}
      \label{Mechanism}
\end{figure}

Some detailed features are designed to improve the performance of the capsule. In Fig. 2(a), a crane-like structure is implemented at the end of the tube, to prevent the capsule from sticking to tissue due to the air pressure generated by vacuum chamber when collecting mucus. Pair of electrodes is also used, as a part of the circuit (see Fig. 3(a)), to make sure that the circuit will close only when mucus exist around the tube. What's more, as in Fig. 2(b) a valve-like structure is also designed to prevent leakage and protect the collected sample.

\subsection{Operation Mechanism}

The operation mechanism of the mucus collection capsule could be described as three main steps. First, before swallowed by patient, the soft chamber will be pressed, and the tube should be sealed with wax, to create and maintain a vacuum environment inside the chamber, as shown in Fig. 3(a). Besides, a permanent magnet will be stuck to the skin of the patient, proximally above the ROI. This magnet only serves as a position marker, in a way of generating strong enough magnetic field within the ROI for triggering the mucus collection process, and have no influence on the capsule's locomotion.

The second step is depicted as Fig. 2(b). The capsule is swallowed by the patient and implements an passive locomotion method propelled by the natural movement of GI tract. After the capsule reaches the desired location, the reed switch will be closed by the magnetic field which is generated by the external magnet on patient's skin, and when mucus exists around the inlet of the chamber at the same time, the circuit will be connected and nichrome wire will start to generate heat to melt the wax. The wax is amorphous material and will melt at around 60 $^\circ$C. The mucus will then be sucked inside the chamber due to the air pressure generated by the chamber's tendency to expand to its original shape. 
The generated heat of nichrome wire will be isolated inside the capsule's soft encapsulate, which is designed to be made of materials with low heat conductivity, and therefore the intestinal tissue will not be damaged by the heat. If cooperate with camera, the condition of sampling site could also be recorded for analysis. 

For the third step, the collected mucus will be protected by capsule and excreted by patient. Further examinations on virus, bacteria or chemical content within the collected mucus could be conducted in laboratory. 

\subsection{Fabrication}
\begin{figure}[!t]
      \centering
      \includegraphics[scale=.45]{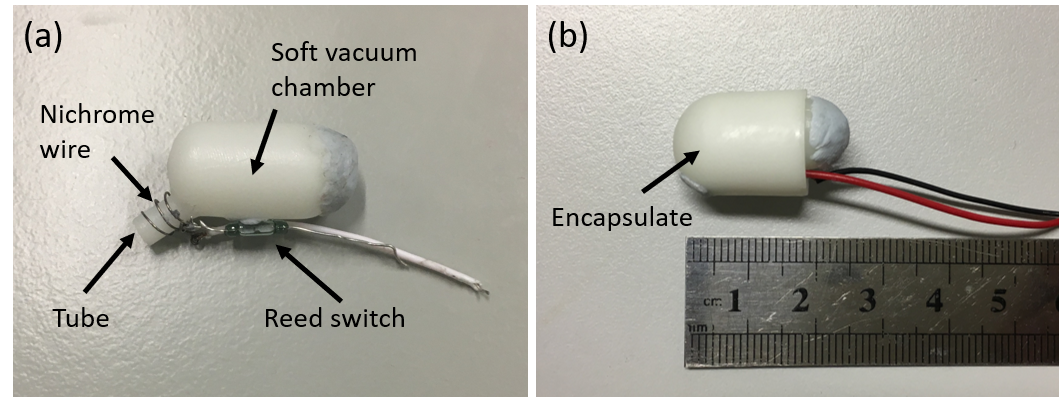}
      \caption{(a) The inner structure of the mucus collection capsule's fabricated prototype for function validation. (b) Encapsulated prototype with a ruler. }
      \label{Prototype}
\end{figure}

The prototype of the soft capsule for function demonstration is shown as Fig. 4(a) and (b). The dimension of the capsule is 14 mm in diameter and 20 mm in length (without camera module and power supply). The soft chamber and encapsulate are fabricated using stereolithography (SLA) 3D printing, with photosensitive silicone as raw material, which has promising heat-resistivity of 70-80 $^\circ$C and satisfying resilience and strength. The hardness of the material is adjustable from 30 to 90 in type A Shore hardness \cite{c23}, and in this design, we choose hardness of 70 to keep a balance between the easiness of deformation and the suction ability. The tube and the chamber are designed as a unitary part to ensure the tightness of the capsule. A hole is opened at the back of the chamber to make it convenient to clean after experiment, and the hole is sealed with reusable adhesive (BluTack, Bostik Inc.) when the capsule is loaded. The encapsulate is designed to cover outside the capsule after all the other parts are assembled.

As for nichrome wire, an alloy with contents of 20\% Ni and 80\% Cr is used, and heat capability of different wires with different lengths and diameters are tested in the experiment. A wire of 40 mm in length and 0.25 mm in diameter is implemented for its proper heat and resistance, and twined into a spiral shape to increase the contact area with the tube. A reed switch (ORD213, Standex-Meder Electronics) with size of 1.8 mm in diameter and 7 mm in length is connected to the nichrome wire. The camera module and the electrodes are not implemented in this prototype and a 1.5 V DC power supply is used as energy source.

\section{Experimental Results and Discussion}

\subsection{Heating Capability of Nichrome Wire} 

\begin{figure}
      \centering
      \includegraphics[scale=.7]{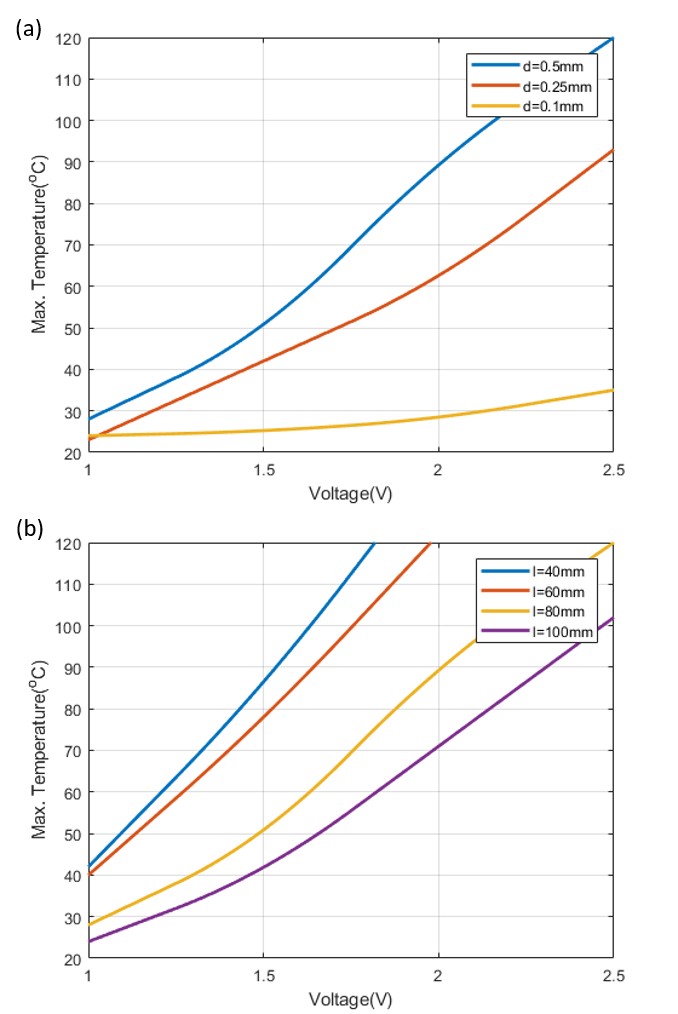}
      \caption{The experimental result for heating capability of nichrome wire. With length of nichrome wires unchanged, the relationship between diameter, applied voltage and maximum temperature is shown in (a). With diameter unchanged, the relationship between wire length, applied voltage and maximum temperature is shown in (b).}
      \label{Heat}
\end{figure}

The heat generation capability of nichrome wire is evaluated in this experiment. Nichrome wires with diameters of 0.1 mm, 0.25 mm and 0.5 mm are tested, considering the dimension of the tube, space in capsule and difficulty of fabrication. Wires of different dimensions are twined into spiral shape and connected to a DC regulated power supply, and temperature is tested using a thermocouple that is connected to a multimeter (UT33, Uni-Trend Technology (China) Co., Ltd.). In the experiment, we define the maximum temperature as the highest temperature that remains unchanged over 10 s after applying current. Besides, the resistances of the nichrome wires are also tested. 

The experiment result is shown as Fig. 5. According to the result, when wire length is unchanged, with diameter increases, the maximum temperature will accordingly increase. With diameter of the wire remains unchanged, smaller length will lead to higher maximum temperature. The resistance of the nichrome wire is as following: 0.275 $\Omega$/cm for diameter 0.5 mm, 0.45 $\Omega$/cm for diameter 0.25 mm, and 0.175 $\Omega$/cm for diameter 0.1 mm. Based on experiment, the nichrome wire implemented in the prototype has a diameter of 0.25 mm and a length of 4 cm, and work under 1.5 V DC power supply which is the same as the voltage provided by conventional batteries for capsule endoscopes. The chosen wire could reach to a maximum temperature of 55 $^\circ$C, at which the wax starts to melt according to real test and the localized heating could be isolated inside the encapsulate of the capsule. 

\subsection{Mucus Collection Test}

\begin{figure}
      \centering
      \includegraphics[scale=.6]{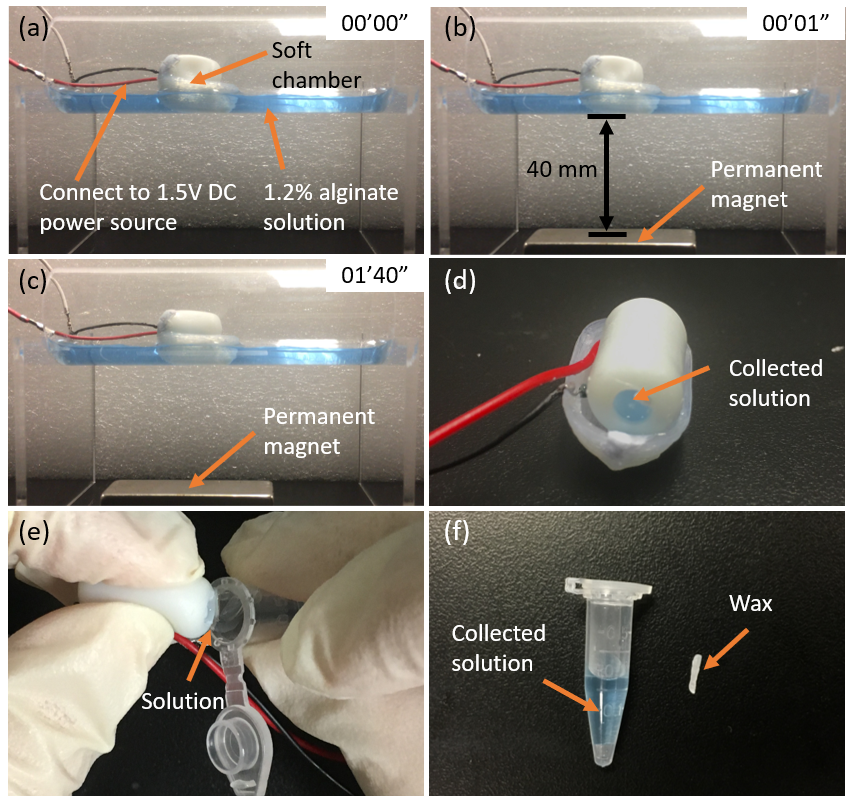}
      \caption{(a) The experiment setup for mucus collection test. (b) At the beginning of the experiment, a permanent magnet was placed 40 mm below the capsule to trigger the sampling process. (c) Collection process finished after 100 s; the chamber expanded to original shape. (d) Solution could be collected in the chamber. (e) Collected solution could be extracted from capsule through the hole on the chamber; (f) 0.3 ml solution was collected and wax was easily separated from sample.}
      \label{Collection}
\end{figure}

In this experiment, the capsule's performance in mucus collection was tested. The experiment setup is shown in Fig. 6(a). A PVC tube was used as the container. An encapsulate that only covers lower half of the chamber was implemented for better view of the soft chamber's state. The capsule was powered by a 1.5 V DC regulated power supply and been merged in 1.2\% alginate solution dyed in blue, which was used to mimic the viscosity of the intestinal mucus while at the same time produce better view of the result. At the start of the experiment, a permanent magnet with surface magnetic field strength of 274 mT was placed at 40 mm below the capsule to trigger the sampling process (see Fig. 6(b)). The 40mm distance is similar to the abdominal subcutaneous fat thickness of obese people \cite{c24}. The minimum magnetic field required to close the reed switch was 48 mT according to measurement. The results are shown in Fig 6(c)-(f). It took approximate 100 seconds for the sampling procedure. The chamber expanded to its original shape and collected solution could be extracted from the chamber through the rear hole. Approximate 0.3 ml solution was collected. The wax was solidified after sampling process and could be easily separated with the collected sample.


Following optimization methods could be carried out to improve the performance of the capsule. The tube, which is made of photosensitive silicone in this design, could be changed to metal to improve the heat conductivity and therefore decrease the respond time and demanded energy. Also, the nichrome wire could be covered with insulation paint to prevent short circuit. Furthermore, the shape and structure of vacuum chamber could be optimized to improve the utilization of space. 

\subsection{Leakage Test in Physiological Condition}
The sealing capability of wax and silicone vacuum chamber is evaluated in this experiment, to ensure that the capsule is able to remain sealed until triggered at the required location. In stomach, the pH value of gastric juice is 1.5-3.5 \cite{c29}, \cite{c17} and the maximum half emptying time is 202 minutes \cite{c18}. In small intestine, the pH value is from 6 to 7.4 \cite{c19} and maximum transit time is 210 minutes \cite{c18}. The leakage test is based on these physiological conditions. 

First, to simulate the environment and transit time inside stomach, which the capsule must pass through using passive locomotion, the sealed chamber was placed in 30 ml HCl solution with pH value equals to 2 for 3.5 hours under the temperature of 37 $^\circ$C. Then, the capsule was moved to 30 ml NaOH solution whose pH value is adjusted to 7.4 and placed for 4 hours. The temperature was also regulated to 37 $^\circ$C. There was no leakage happened after the 7.5 hours' test and no obvious flaw observed on the surface of chamber and encapsulate. Experimental conditions and results are shown in Table 1.


\renewcommand{\arraystretch}{1.2}
\begin{table}
\caption{Conditions and Results of Reliability Test}
\label{table_example}
\centering
\begin{center}
\begin{tabular}{p{1.5cm} p{1.5cm} p{1.2cm} c}
\hline
\hline
  \centering & \multicolumn{2}{c}{ Conditions} & Results\\
\hline
   \centering ~ & Temperature & 37 $^\circ$C & ~ \\
   \centering Stomach & pH value & 2 & No leakage \\
   \centering ~ & Time & 3.5 h & ~ \\
\hline 
   \centering ~ & Temperature & 37 $^\circ$C & ~ \\
   \centering Intestine & pH value & 7.4 & No leakage \\
   \centering ~ & Time & 4 h & ~ \\
\hline
\hline
\end{tabular}
\end{center}
\end{table}

\section{Conclusions and Future Work}

In this work, a soft capsule for mucus collection is designed and a prototype for function demonstration is fabricated. The capsule implements a soft chamber which is sealed with wax to generate vacuum environment for mucus suction, and moves passively along the GI tract. The mucus collection procedure is triggered by a permanent magnet that sticks on patient's skin proximally above the ROI. Experiments are conducted to show the validity of the design. 


Future work will mainly focus on the following aspects to further complete the results and enhance the efficiency and stability of the capsule. 

\begin{itemize}

\item Optimization on the capsule design should be made in the next prototypes to shorten the response time and increase the mucus collection capability, including improving the heat conductivity of the tube, changing the hardness of the chamber, etc. 
\item Mucus collection capability of the capsule should be further tested under various experimental conditions, for example, on soft tissue surface and in the environment with thinner mucus layer.
\item Rolling and dragging movements should be added to the leakage test of further prototypes. 
\item The electrical conductivity of mucus will be tested through ex-vivo experiments in further prototypes.
\item In-vivo experiment could be carried out when this technique is more refined.

\end{itemize}

Furthermore, extension of function will also be considered in the future. When combines with bacteria sensing technologies \cite{c20}, the capsule may achieve in-vivo mucus analysis in further prototypes. Cargo delivery function may also be designed for releasing swarming microrobots in specific location \cite{c21}-\cite{c28} for targeted therapy or other biomedical applications. 

\addtolength{\textheight}{-4cm}   




\section*{Acknowledgment}

This research is financially supported by Innovation and Technology Fund (project number: ITS/440/17FP), CUHK T Stone Robotics Institute and CUHK-SJTU Joint Research Center on Medical Robotics, The Chinese University of Hong Kong. The authors would like to express their sincere gratitude to all the members in Professor Li Zhang's research group at CUHK for their valuable suggestions.


\end{document}